\documentclass[conference]{IEEEtran}

\usepackage{cite}
\usepackage[cmex10]{amsmath}

\usepackage{tabularx,colortbl}
\usepackage{amsmath,amssymb,amsthm,mathrsfs,dsfont}

\usepackage{amsmath}
\usepackage{amssymb}

\usepackage{threeparttable}
\usepackage{multirow}

\usepackage{algorithm}
\usepackage{algpseudocode}

\usepackage{subfigure}
\usepackage{stfloats}

\newcounter{MYtempeqncnt}
\IEEEoverridecommandlockouts
\ifCLASSINFOpdf
  \usepackage[pdftex]{graphicx}
  \DeclareGraphicsExtensions{.pdf,.jpeg,.png}
\else
  \usepackage[dvips]{graphicx}
  \DeclareGraphicsExtensions{.eps}
\fi

\begin{document}

\title{\huge{Wireless Energy Harvesting Cooperative Communications with Direct Link and Energy Accumulation}
\thanks{This research was supported by ARC grants DP150104019 and FT120100487. The research was also supported by funding from the Faculty of Engineering and Information Technologies, The University of Sydney, under the Faculty Research Cluster Program and the Faculty Early Career Researcher Scheme.}}

\author{\IEEEauthorblockN{Ziyi Li, He Chen, Yonghui Li, and Branka Vucetic}
\IEEEauthorblockA{School of Electrical and Information Engineering,
University of Sydney, Sydney, NSW 2006, Australia\\
E-mail:~\{ziyi.li,~he.chen,~yonghui.li,~branka.vucetic\}@sydney.edu.au
}
}

\maketitle
\begin{abstract}
This paper investigates a wireless energy harvesting cooperative network (WEHCN) consisting of a source, a decode-and-forward (DF) relay and a destination. We consider the relay as an energy harvesting (EH) node equipped with EH circuit and a rechargeable battery. Moreover, the direct link between source and destination is assumed to exist. The relay can thus harvest and accumulate energy from radio-frequency signals ejected by the source and assist its information transmission opportunistically. We develop an incremental accumulate-then-forward (IATF) relaying protocol for the considered WEHCN. In the IATF protocol, the source sends its information to destination via the direct link and requests the relay to cooperate only when it is necessary such that the relay has more chances to accumulate the harvested energy.
By modeling the charging/discharging behaviors of the relay battery as a finite-state Markov chain, we derive a closed-form expression for the outage probability of the proposed IATF. Numerical results validate our theoretical analysis and show that the IATF scheme can significantly outperform the direct transmission scheme without cooperation.


\end{abstract}
\IEEEpeerreviewmaketitle

\section{Introduction}
Wireless energy harvesting (WEH) technique has recently emerged as a promising solution to extend the lifetime of energy constrained wireless networks, e.g., wireless sensor networks (see \cite{xiao2014wireless,chen2015harvest} and references therein).
WEH technique has also brought new research opportunities for cooperative communications as it enables a new cooperative manner for wireless devices. Specifically, the relay node is now able to harvest energy from the source's information and then use the harvested energy to assist its transmission. In this sense, the relay is more willing to cooperate with the source since it does not need to consume its own energy. In this paper, we refer to a cooperative communication network with a WEH relay as a wireless energy harvesting cooperative network (WEHCN). In fact, the design and analysis of WEHCNs have attracted considerable attentions recently (see, e.g., \cite{krikidis2012rf,nasir2015block,krikidis2015relay,liu2015performance,gu2016distributed}).

Due to the propagation loss of RF signals, an energy harvesting (EH) node normally harvests a little amount of energy during each scheduling slot, which may lead to poor network performance if the node exhausts this little amount of energy to perform instantaneous information transmission. Thus, it is essential to equip each EH node with an energy storage (e.g., a rechargeable battery) such that they can accumulate sufficient amount of harvested energy and then transmit its information in an appropriate time slot. However, to our best knowledge, only a few papers in open literature incorporated the energy accumulation (EA) process in the design and analysis of WEHCNs\footnote{Note that the EA process has also been studied for wireless networks with EH from natural resources (e.g., solar). However, in such networks, the EH process is in fact independent of the network operations. As such, the modeling and analysis therein cannot be used for the considered WEHCNs.}. The EA process of three-node cooperative networks with a single EH relay was modeled and the associated network performance was analyzed in \cite{krikidis2012rf,nasir2015block}. Very recently, \cite{krikidis2015relay,liu2015performance,gu2016distributed} investigated multi-relay WEHCNs, in which different relay selection schemes were proposed and analyzed by taking into consideration the EA process at relay nodes.

In \cite{krikidis2012rf,nasir2015block,krikidis2015relay,liu2015performance,gu2016distributed}, it is assumed that the direct link between source and destination is not existent for simplicity and the communication between them completely relies on the EH relay(s). Nevertheless, the direct link may play a crucial role in WEHCNs as it could substantially contribute to the received signal-to-noise ratio (SNR) at the destination. This is due to the fact that the harvested energy at relay(s) is normally small and its transmit power would be limited.
The operation of WEHCNs with a direct link is not so straightforward as those without direct link. Specifically, without direct link, the source fully depends on the EH relay, which should accumulate enough energy before helping forward source's information to destination. In contrast, when the direct link exists, the source is able to deliver information directly to the destination and charge the relay wirelessly at the same time. This indicates that the EA process of the EH relay depends on not only its own status but also the channel condition of the direct link, which makes the design and analysis of the considered WEHCN non-trivial.
To the best knowledge of authors, there is no work in open literature that designs and analyzes an WEHCN with EA at the EH relay and a direct link between source and destination.

Motivated by this gap, in this paper we investigate a typical three-node WEHCN consisting of one source, one EH relay with multiple antennas, and one destination. We consider that a direct link exists between source and destination. The \emph{main contributions} of this paper are summarized as follows: \textbf{(1)} To efficiently utilize the EH relay to improve the system performance, we develop an incremental accumulate-then-forward (IATF) relaying protocol for the considered WEHCN.
Inspired by conventional incremental relaying protocol \cite{laneman2004cooperative,Chen_WCSP_2009}, the proposed IATF protocol advocates that the destination first relies on the received signal via direct link to decode source's information and requests the relay to forward source's signal only when an outage occurs in the direct link. As such, the relay has more chances to accumulate energy from source's signals and help the source effectively when it is necessary.
\textbf{(2)} We model the dynamic behaviors of relay battery as a finite-state Markov chain and then evaluate the discrete distribution of the relay. With the help of the steady state distribution of the relay residual energy, we then derive a closed-form expression for the outage probability of the proposed IATF protocol over mixed Rician-Rayleigh fading channels. \textbf{(3)} Numerical results show that the proposed IATF protocol can outperform the direct transmission scheme without the help of relay and the performance gain is enlarged when either the source transmit power or the number of antennas at relay increases.
\section{System Model}\label{System}
We consider a WEHCN consisting of one single-antenna source $S$, one single-antenna destination $D$ and one decode-and-forward (DF) relay $R$ equipped with $N$ antennas ($N\geq1$), which is deployed to assist the data transmission from $S$ to $D$. We assume $S$ and $D$ are connected with external power supply, while $R$ has no embedded power supplies. But, $R$ is equipped with an EH unit and a finite-size rechargeable battery such that it can accumulate the energy harvested from RF signals broadcasted by $S$. Moreover, the relay is equipped with separate energy and information receivers \cite{zhang2013mimo}. As such, it can flexibly switch its received signal to one of these two receivers to realize EH or information decoding (ID). Furthermore, different from the existing works on WEHCNs, we consider the scenario that the direct link between $S$ and $D$ exists.

Let $h_{SD}$ denote the complex channel coefficient of $S$-$D$ link. Also let $\mathbf{h}_{SR}$ and $\mathbf{h}_{RD}$ denote the $N\times1$ channel vectors of $S$-$R$ and $R$-$D$ link, respectively. As the up-to-date WEH techniques could only be operated within a relatively short distance, the line-of-sight (LoS) path is very likely to exist between $S$ and $R$. Motivated by this fact, we consider an asymmetric scenario for the fading distributions of $S$-$R$ link and $R$-$D$ link. Specifically, the elements of $\mathbf{h}_{SR}$ are subject to independent and identically distributed (i.i.d.) Rician fading, while the elements of $\mathbf{h}_{RD}$ are subject to i.i.d. Rayleigh fading as the distance between $R$ and $D$ can be much further. Similarly, the channel coefficient of direct link $h_{SD}$ follows a Rayleigh distribution. All channels between $S$, $R$ and $D$ are assumed to experience slow, independent and frequency flat fading such that the channel gains remain unchanged within each transmission block but change independently from one block to the other.

Now, a natural question that arises is ``With the direct link, how to optimally utilize the EH relay to minimize the system outage probability?". This is actually a very difficult question to answer by realizing the fact that the network operation and the EA process of the EH relay are tangled together. Motivated by this, in this paper we propose a heuristic cooperative protocol, termed incremental accumulate-then-forward (IATF), to efficiently utilized the EH relay in the considered WEHCN. In the proposed IATF protocol, the EH relay cooperates with $S$ opportunistically using the accumulated energy harvested from signals broadcasted by $S$.
Here, we use $T$ to denote the duration of each transmission block, which is further divided into two consecutive time slots with equal length $T/2$. In the first time slot, $S$ first broadcasts the information and meanwhile $R$ decides whether to perform EH or ID based on its accumulated energy available in battery. After $D$ decodes the received signal from $S$ at the end of the first time slot, it feedbacks a one-bit signal to both $S$ and $R$, indicating whether the direct link suffers from an outage event. In the second time slot, the cooperation between $S$ and $R$ will be activated if the $S$-$D$ link fails and the residual energy at $R$ is not less than a predefined energy threshold $E_T$. Otherwise, during the second time slot, {$S$ chooses to either re-transmit the same information packet or send a new one depending on whether the $S$-$D$ link fails or not, while $R$ always harvests energy from the signal sent by $S$}. Let $X$ denote the direct link status indicator, which equals to $0$ if it suffers from no outage and $1$ otherwise, and $Y$ denote the energy status indicator at $R$, which equals to $0$ if the available energy at $R$ is less than $E_T$ and $1$ otherwise. Depending on the values of indicators $X$ and $Y$, the system can operate in one of the following four modes.
\begin{itemize}
\item Mode I ($X = 0$ and $Y = 0$): In this mode, in the first time slot, $R$ harvests energy since the residual energy at its battery is less than $E_T$. Also, in the second time slot, there is no need for $R$ to forward information since the $S$-$D$ link is outage-free. As such, in the second time slot, $S$ will send new information to $D$, and $R$ can further accumulate its energy harvested from this signal.
\item Mode II ($X = 0$ and $Y = 1$): Here, in the first time slot, $R$ chooses to decode the received information since it has accumulated sufficient energy and the direct link suffers from no outage. Thus, the operation of the system in the second time slot is the same as that in Mode I.
\item Mode III ($X = 1$ and $Y = 0$): $R$ harvests energy during the first time slot in this mode. $X=1$ indicates the direct transmission fails and in this case it needs $R$ to help forward the information to $D$. However, $R$ cannot help since its accumulated energy is lower than the threshold $E_T$. In this case, $S$ chooses to further charge $R$ by re-transmitting its information packet during the second time~slot.
\item Mode IV ($X = 1$ and $Y = 1$): In this mode, the direct link suffers from an outage event and $R$ has sufficient energy to cooperate. Thus, in the second time slot, $R$ forwards the signal received from $S$ to $D$ by consuming $E_T$ amount of energy from its battery, while $S$ keeps silent during this period.

\end{itemize}

We now analyze the harvested energy at $R$ and the received SNR at $D$ for the above four modes mathematically in the following subsections. Without loss of generality, we consider a normalized transmission block (i.e., $T=1$) hereafter. Besides, we use $P_S$ to denote the transmit power at $S$ and $H_{x,y}=\|\mathbf{h}_{x,y}\|^2$ to denote the channel power gain, where $x \in \{S, R\}$, $y \in \{D\}$ and $\|\mathbf{x}\|$ denotes the Euclidean norm of a vector $\mathbf{x}$.
\subsection{Mode I}
In Mode I, the direct link has a good channel condition. After $D$ correctly decodes the information in the first time slot, $S$ chooses to continuously transmit new information. The received SNR at $D$ in both time slots is given by
\begin{equation}\label{SD}
\gamma_D^{\rm{I}} = \gamma_{SD}={P_S H_{SD}}/{N_0},
\end{equation}
where $N_0$ is the power of the additive Gaussian white noise~(AWGN).

Recall that in this mode the residual energy at $R$ is less than the energy threshold and the $S$-$D$ link is outage-free. Thus $R$ harvests energy in both time slots. By ignoring the amount of energy harvested from the noise power since it is normally below the sensitivity of WEH units, the amount of harvested energy at $R$ in Mode I is expressed as
\begin{equation}
E_H^{\rm{I}} = E_{H1} = \eta P_S H_{SR},
\end{equation}
where $\eta \in{\left(0,1\right]}$ is the energy conversion efficiency.

\subsection{Mode II}
In this mode, $R$ decodes the received information since the residual energy in its battery exceeds $E_T$. Meanwhile, there is no outage between $S$ and $D$. As such, in the second time slot, $S$ transmits new information.
We assume that the maximum ratio combining (MRC) technique is adopted at $R$ to maximize the received SNR. In this case, the received SNR at $R$ in the first time slot is given by
\begin{equation}
\gamma_{SR} = {P_S H_{SR}}/{N_0}.
\end{equation}
And the received SNR at $D$ is the same as that in Mode I (i.e., $\gamma_D^{\rm{II}}=\gamma_D^{\rm{I}}$).
Besides, in the second time slot, the relay can harvest energy from the new packet transmitted by $S$. The harvested energy at $R$ in this mode can thus be expressed~as
\begin{equation}
E_{H}^{\rm{II}} = E_{H2} ={\eta P_S H_{SR}}/{2}.
\end{equation}

\subsection{Mode III}
In Mode III, the channel condition of direct link is in a bad condition, which causes a transmission outage. In this case, the system needs relay's cooperation to forward source's signal to $D$. However, the residual energy at relay battery is less than $E_T$. $S$ is motivated to charge $R$ by re-transmitting its information packet in the second time slot. The received SNR at $D$ is thus given by
\begin{equation}
\gamma_D^{\rm{III}} = 2 \gamma_{SD} = {2 P_S H_{SD}}/{N_0}
\end{equation}
And $R$ harvests energy in both time slots and the harvested energy $E_{H}^{\rm{III}} = E_{H1}$.

\subsection{Mode IV}
In this mode, the $S$-$D$ transmission fails and $R$ has accumulated enough energy. Thus, $R$ will forward the source's signal in the second time slot by consuming the amount of energy $E_T$. We use $P_R$ to denote the relay transmit power and we thus have $P_R = 2E_T$ since the relay consumes $E_T$ within half of a transmission block. We assume that before the cooperation, $D$ sends a pilot signal to $R$ such that the CSI $\mathbf{h}_{RD}$ of the corresponding reciprocal channel can be estimated. In this case, maximum ratio transmission (MRT) is known to be optimal for data transmission at $R$. The received SNR at $D$ in the second time slot is thus given by
\begin{equation}
\gamma_{RD} = {P_R H_{RD}}/{N_0},
\end{equation}
Note that the received SNR at $D$ during the first time slot is the same as (\ref{SD}). $D$ then combines these two signals using MRC technique.
With reference to \cite[Eqs. (16)-(17)]{laneman2004cooperative}, the received SNR at $D$ in Mode IV can be expressed as
\begin{equation}\label{gamma_3}
\gamma_{D}^{\rm{IV}} = \min{\left(\gamma_{SR},\gamma_{SD}+\gamma_{RD}\right)}.
\end{equation}

\section{{Outage Probability Analysis}}\label{Analysis}
In this section, we analyze the system outage probability of the proposed IATF protocol over mixed Rician-Rayleigh fading channels. To this end, we first model the dynamic behaviors of the relay battery as a finite-state Markov chain (MC)~\cite{krikidis2012rf}.

\subsection{Markov Chain Description of Relay Battery}
We follow \cite{krikidis2012rf} and assume that $R$ is equipped with a $L$ discrete-level battery with a finite capacity $C$. The $i$th energy level is then defined as $\varepsilon_i = i C / L$, $i\in \{0,1,2,\ldots,L \}$. We define the state $S_i$ as the state of relay's residual battery being $\varepsilon_i$. $P_{i,j}$ is defined as the state transition probability from $S_i$ to $S_j$. Let $\lambda[m]\in\{\lambda_{{\rm{I}}},\lambda_{{\rm{II}}},\lambda_{{\rm{III}}},\lambda_{{\rm{IV}}}\}$ denote the system operation mode in the $m$-th transmission block, where $\lambda_M$, $M\in{\{{\rm{I}},{\rm{II}},{\rm{III}},{\rm{IV}}\}}$, represents the event that the Mode $M$ is operated. With the discrete battery model adopted in this paper, the effective (discretized) amount of harvested energy $\varepsilon_H^M$ at $R$ should be re-calculated as
\begin{equation}\label{varepsilon_H}
\small{
\varepsilon_H^M \triangleq \varepsilon_i,~{\rm{where}}~ i = \arg\max\nolimits_{j\in\{0,1,\ldots,L\}}\left\{\varepsilon_j:\varepsilon_j < E_H^M\right\},
}\end{equation}
where $M\in\{{\rm{I}},{\rm{II}},{\rm{III}}\}$. Similarly, the actual amount of energy consumed by $R$ for information forwarding should be defined~by
\begin{equation}\label{varepsilon_T}
\varepsilon_T \triangleq \varepsilon_i,~{\rm{where}}~ i = \arg\min\nolimits_{j\in\{1,\ldots,L\}}\left\{\varepsilon_j:\varepsilon_j\geq E_T\right\}.
\end{equation}
In this paper, we define that an outage occurs in a link when the system transmission rate is larger than the link capacity.
Let $\mathbb{R}$ denote the transmission rate of $S$. Then the direct link suffers from an outage when the received SNR $\gamma_{SD}$ is less than the SNR threshold $\gamma_1 = 2^{\mathbb{R}}-1$ \cite{laneman2004cooperative}.
We now can describe the four possible operations of the proposed IATF protocol during the $m$-th transmission block mathematically as follows:
\begin{equation}
\small
\lambda[m] =
\begin{cases}
\lambda_{\rm{I}}, &\mbox{if ~$\gamma_{SD}\geq\gamma_1~\&~\varepsilon[m]<\varepsilon_T$},\\
\lambda_{\rm{II}}, &\mbox{if ~$\gamma_{SD}\geq\gamma_1~\&~\varepsilon[m]\geq\varepsilon_T$},\\
\lambda_{\rm{III}}, &\mbox{if ~$\gamma_{SD}<\gamma_1~\&~\varepsilon[m]<\varepsilon_T$},\\
\lambda_{\rm{IV}}, &\mbox{if ~$\gamma_{SD}<\gamma_1~\&~\varepsilon[m]\geq\varepsilon_T$},\\
\end{cases}
\end{equation}
where $\varepsilon[m]$ denotes the relay's residual energy at the beginning of the $m$-th transmission block. Moreover, the residual energy at the beginning of the $(m+1)$th transmission block can thus be expressed as
\begin{equation}
\small
\varepsilon[m+1] =
\begin{cases}
\min \{\varepsilon[m] + \varepsilon_H^{\rm{I}}, C \}, &\mbox{if $\lambda[m] = \lambda_{\rm{I}}$}\\
\min \{\varepsilon[m] + \varepsilon_H^{\rm{II}}, C \}, &\mbox{if $\lambda[m] = \lambda_{\rm{II}}$}\\
\min \{\varepsilon[m] + \varepsilon_H^{\rm{III}}, C \}, &\mbox{if $\lambda[m] = \lambda_{\rm{III}}$}\\
\varepsilon[m] - \varepsilon_T, &\mbox{if $\lambda[m] = \lambda_{\rm{IV}}$}\\
\end{cases}.
\end{equation}

Based on the above mathematical description, we now derive the state transition probabilities of the formulated MC for the relay battery. Inspired by \cite{krikidis2012rf}, the state transition of the MC can be generally split into the following eight cases.
%

%
\subsubsection{The battery remains empty ($S_0$ to $S_0$)}
We consider the initial state of the MC starts from $S_0$. It is easy to deduce that $R$ will operate in either Mode I or Mode III. Furthermore, the effective harvested energy during the whole transmission block should be zero, which indicates that the condition $E_{H1}<C/L$ should hold. Thus, the transition probability of this case is characterized as
\begin{equation}
\small
\begin{split}
P_{0,0} &= \Pr\left\{E_{H1} < C/L \right\} =\Pr\left\{H_{SR}<C/\left(\eta P_S L\right)\right\}\\
&= F_{H_{SR}}\left(\frac{C}{\eta P_S L}\right),
\end{split}
\end{equation}
where $F_{H_{SR}}\left(\cdot\right)$ denotes the cumulative distribution function (CDF) of $H_{SR}$. According to \cite{ko2000average}, we can write the CDF of $H_{SR}$ as
$\small {F_{H_{SR}}\left(x\right) = 1-Q_N\left(\sqrt{2NK},\sqrt{\frac{2(K+1)}{\Omega_{SR}}x}\right)},$
where $Q_N\left(\cdot,\cdot\right)$ is the generalized ($N$th-order) Marcum $Q$-function \cite{zwillinger2014table}, $K$ is the Rician $K$-factor defined as the ratio of the powers of the LoS component to the scattered components and $\Omega_{SR} = \mathbb{E}\{\left|h_{SR,i}\right|^2\},~\forall i\in\left\{1,\ldots,N\right\}$, with $\mathbb{E}\{\cdot\} $ denoting the statistical expectation and $h_{SR,i}$ denoting the $i$-th element of ${\bf h}_{SR}$.

\subsubsection{The empty battery is partially charged ($S_0$ to $S_i$ with $0<i<L$)}
In this case, the amount of effective energy is discretized as $\varepsilon_H^M=iC/L, M \in {\{{\rm{I}},{\rm{III}}\}}$. This indicates $E_{H1}$ falls between the battery levels $i$ and $i+1$. The transition probability can thus be evaluated as
\begin{equation}
\small
\begin{split}
P_{0,i} &= \Pr\left\{iC/L \leq E_{H1} < \left(i+1\right)C/L\right\} \\
        &= F_{H_{SR}}\left(\frac{\left(i+1\right)C}{\eta P_S L}\right) -  F_{H_{SR}}\left(\frac{i C}{\eta P_S L}\right).
\end{split}
\end{equation}

\subsubsection{The empty battery is fully charged ($S_0$ to $S_L$)}
Similar to the previous two cases, the transition probability is characterized~as
\begin{equation}
\small
P_{0,L} = \Pr\left\{E_{H1} \geq C \right\} = 1 - F_{H_{SR}}\left(\frac{C}{\eta P_S}\right).
\end{equation}

\subsubsection{The non-empty and non-full battery remains unchanged ($S_i$ to $S_i$ with $0<i<L$)}\label{ii_section}
The battery level stays unchanged, from which we can deduce that the $R$ is possible to operate in Mode I, II or III with effective harvested energy equals to zero (i.e., $E_H^M$ is discretized into zero). The transition probability of this case is derived as
\begin{equation}\label{ii}
\small{
\begin{split}
&{P_{i,i}}= \Pr \left\{ \left[ \left(Y=0\right) \cap \left( {E_{H1} < \varepsilon_1} \right) \right] \right.\\
 &~~~~~~~~~~\left.\cup \left[ \left(X=0\right) \cap \left(Y=1\right) \cap \left( {E_{H2} < \varepsilon_1} \right) \right]\right\}\\
        &=
        \begin{cases}
F_{H_{SR}}\left( {\frac{C}{{\eta {P_S}L}}} \right),&~\mbox{if $E_T > \frac{i C}{L}$};\\
\left[1-F_{H_{SD}}\left(\frac{\gamma_1 N_0}{P_S}\right)\right]F_{H_{SR}}\left( {\frac{2 C}{{\eta {P_S}L}}} \right),&~\mbox{if $E_T \leq \frac{i C}{L}$},\\
        \end{cases}
\end{split}
}
\end{equation}
where $F_{H_{SD}}\left(x\right)=1-\exp\left(-x/\Omega_{SD}\right)$ with $\Omega_{SD} = \mathbb{E}\{\left|h_{SD}\right|^2\}$ denoting the mean of $H_{SD}$.

\subsubsection{The non-empty battery is partially charged ($S_i$ to $S_j$ with $0<i<j<L$)}
Similar with the previous case, the battery level increases from level $i$ to $j$ (i.e., $\varepsilon_H^{M} = \left(j-i\right)C/L$). Thus the transition probability can be calculated as
\begin{equation}\label{ij}
\small{
\begin{split}
&P_{i,j} = \Pr \left\{ \left[ \left(Y=0\right) \cap \left( \varepsilon_{j-i} \leq E_{H1} < \varepsilon_{j-i+1} \right) \right]\right.\\
&~~~~~~~~~\left.\cup \left[ \left(X=0\right) \cap \left(Y=1\right) \cap \left( \varepsilon_{j-i} \leq E_{H2} < \varepsilon_{j-i+1} \right) \right] \right\}\\
        &=
        \begin{cases}
        F_{H_{SR}}\left(\frac{\left(j-i+1\right)C}{\eta P_S L}\right)-F_{H_{SR}}\left(\frac{\left(j-i\right)C}{\eta P_S L}\right),\mbox{if $E_T > \frac{i C}{L}$};            \\
        \left[F_{H_{SR}}\left(\frac{2\left(j-i+1\right)C}{\eta P_S L}\right)-F_{H_{SR}}\left(\frac{2\left(j-i\right)C}{\eta P_S L}\right)\right]\times\\
        \left[1-F_{H_{SD}}\left(\frac{\gamma_1 N_0}{P_S}\right)\right],~~~~~~~~~~~~~~~~~~~\mbox{if $E_T \leq \frac{i C}{L}$}.\\
        \end{cases}
\end{split}
}\end{equation}

\subsubsection{The non-empty and non-full battery is fully charged ($S_i$ to $S_L$ with $0<i<L$)}
In this case, the harvested energy $E_H^M$ can be any value greater than $\left(L-i\right)C/L$, the transition probability can thus be evaluated by
\begin{equation}\label{iL}
\small{
\begin{split}
&{P_{i,L}} = \Pr \left\{ \left[ {\left(Y=0\right) \cap \left( {E_{H1} \geq \varepsilon_{L-i}} \right)} \right] \right.\\
&~~~~~~~~~~~\left.\cup \left[ \left(X=0\right) \cap \left(Y=1\right) \cap \left( {E_{H2} \geq \varepsilon_{L-i}} \right) \right]\right\}\\
        &=
        \begin{cases}
        1-F_{H_{SR}}\left( {\frac{\left(L-i\right)C}{{\eta {P_S}L}}} \right),~~~~~~~~~~~~~~~~~~~~~~~~~~~~~\mbox{if $E_T > \frac{i C}{L}$};\\
        \left[1-F_{H_{SD}}\left(\frac{\gamma_1 N_0}{P_S}\right)\right]\left[1-F_{H_{SR}}\left( {\frac{2\left(L-i\right)C}{{\eta {P_S}L}}}\right)\right],~\mbox{if $E_T \leq \frac{i C}{L}$}.
        \end{cases}
\end{split}}
\end{equation}

\subsubsection{The battery remains full ($S_L$ to $S_L$)}
In this case, Mode II is definitely performed with a full battery and the help of $R$ is not necessary so that the battery level remains unchanged. Recall the principle of the proposed IATF scheme, the transition probability is derived as
\begin{equation}\label{LL}
\small
P_{L,L} = \Pr\left\{X=0\right\}= 1-F_{H_{SD}}\left(\frac{\gamma_1 N_0}{P_S}\right).
\end{equation}

\subsubsection{The non-empty battery discharged ($S_j$ to $S_i$ with $0\leq i\leq j\leq L$)}
The battery level decreases only when Mode IV is operated in the system. The transition probability can thus be characterized as
\begin{equation}\label{ji}
\small{
\begin{split}
P_{j,i} &= \Pr\left\{\left(X=1\right) \cap \left(Y=1\right) \cap \left(E_T = \left(j-i\right)C/L\right)\right\}\\
        &=
        \begin{cases}
        F_{H_{SD}}\left(\frac{\gamma_1 N_0}{P_S}\right),
        &\mbox{if $E_T = \frac{\left(j-i\right)C}{L}$};        \\
        0,
        &\mbox{if $E_T \neq \frac{\left(j-i\right)C}{L}$}.        \\
        \end{cases}
\end{split}
}
\end{equation}

We are now ready to derive the steady state distribution of the relay battery. Let $\mathbf{Z} = \left[P_{i,j}\right]_{(L+1)\times(L+1)}$ denote the state transition matrix of the formulated MC. Using similar method used in \cite{krikidis2012buffer}, we can readily verify that $\mathbf{Z}$ is irreducible and row stochastic. Thus, there should exists a unique solution $\boldsymbol{\pi}$ that satisfies the following equation
\begin{equation}
\mathbf{\boldsymbol{\pi}} = \left(\pi_0,\pi_1,\ldots,\pi_L\right)^T=\mathbf{Z}^T \boldsymbol{\pi}.
\end{equation}
This $\boldsymbol{\pi}$ is actually the steady state distribution of the relay residual energy and can be calculated as
\begin{equation}
\boldsymbol{\pi} = \left(\mathbf{Z}^T - \mathbf{I} + \mathbf{B}\right)^{-1}\mathbf{b},
\end{equation}
where $\mathbf{Z}^T$ denotes the transpose matrix of $\mathbf{Z}$, $\mathbf{I}$ is the identity matrix, $B_{i,j} = 1, \forall{i,j}$, and $\mathbf{b} = \left(1,1,\ldots,1\right)^T$\cite{krikidis2012buffer}.

\subsection{System Outage Probability}
We now analyze the system outage probability of the proposed IATF scheme based on the steady state of the relay battery derived in the previous subsection. Let $P_{\rm{out}}^M$, $M\in\{{\rm{I}}, {\rm{II}}, {\rm{III}},{\rm{IV}}\}$, denote the probability that an outage occurs when the system operates in Mode I, II, III and IV, respectively. According to the full probability theory, we can express the outage probability of the considered WPCCN as
\begin{equation}\label{P_out}
P_{\textrm{out}} = \sum_{X\in\{\textrm{I},\textrm{II},\textrm{III},\textrm{IV}\}} \Pr\left\{M\right\} \Pr\left\{O|M\right\},
\end{equation}
where $\Pr\left\{M\right\}$ is the probability that the system operates in Mode $M$, and $\Pr\left\{O|M\right\}$ is the system outage probability when it operates in Mode $M$.

In Mode I and Mode II, no outage event happens. We thus have $\Pr\left\{O|\textrm{I}\right\}=\Pr\left\{O|\textrm{II}\right\}=0$ and do not need to calculate the probabilities $\Pr\left\{\textrm{I}\right\}$ and $\Pr\left\{\textrm{II}\right\}$.

In Mode III and IV, the system delivers the same information packet in two time slots. In this case, the spectrum efficiency is halved. As such, the outage occurs in the system when $\gamma_D^M$, $M\in\{{\textrm{III}},{\textrm{IV}}\}$, is less than a higher threshold $\gamma_2 = 2^{2\mathbb{R}}-1$ \cite{laneman2004cooperative}. In Mode III, the direct link is in outage and $D$ receives two pieces of the same information from $S$. The probability of the system operates in Mode III can be expressed as
\begin{equation}\label{P3}
\begin{split}
P\left\{O|\textrm{III}\right\} &= \left(1-P_E\right)\Pr\left\{\gamma_{SD}<\gamma_1\right\}\\
&=\left(1-P_E\right) F_{H_{SD}}\left(\frac{\gamma_1 N_0}{P_S}\right),
\end{split}
\end{equation}
where $P_E$ denotes the probability that $R$ has enough residual energy to cooperate, which can be characterized based on the derived steady state of relay battery as
\begin{equation}\label{P_E}
P_E = \sum\nolimits_{i=k}^{L} \pi_i,~{\textrm{s.t.}}~k = \arg\min_{k\in1,\ldots,L}\left\{\varepsilon_k \geq E_T\right\}.
\end{equation}
Besides, it can be readily verified that the outage probability of Mode III $\Pr\left\{O|\textrm{III}\right\} = \Pr\left\{\gamma_D^{\textrm{III}}<\gamma_2|\gamma_{SD}<\gamma_1\right\}=1$ for any $\mathbb{R} > 0$.

Similarly, the product of the probability of Mode IV and its associated outage probability can be derived as
\begin{equation}\label{P4_1}
\begin{split}
&\Pr\left\{\textrm{IV}\right\} \Pr\left\{O|\textrm{IV}\right\} \\
&= P_E \Pr\left\{\gamma_{SD}<\gamma_1\right\}\Pr\left\{\gamma_D^{\textrm{IV}}<\gamma_2|\gamma_{SD}<\gamma_1\right\}\\
&=P_E\Pr\left\{\gamma_{SR}\geq\gamma_2\right\}\Pr\left\{\left(\gamma_{SD}+\gamma_{RD}<\gamma_2\right)\cap\left(\gamma_{SD}<\gamma_1\right)\right\}\\
&~~+P_E \Pr\left\{\gamma_{SD}<\gamma_1\right\} \Pr\left\{\gamma_{SR}<\gamma_2\right\},
\end{split}
\end{equation}
where $\Pr\{\gamma_{SR}\geq\gamma_2\}$ $=$ $1-F_{H_{SR}}\left(\gamma_2 N_0/P_S\right)$, $\Pr\{\gamma_{SD}<\gamma_1\}$ $=$ $F_{H_{SD}}\left(\gamma_1 N_0/P_S\right)$ and $\Pr\{\gamma_{SR}<\gamma_2\}$ $=$ $F_{H_{SR}}\left(\gamma_2 N_0/P_S\right)$. Moreover, the term $\Pr\left\{\left(\gamma_{SD}+\gamma_{RD}<\gamma_2\right)\cap\left(\gamma_{SD}<\gamma_1\right)\right\}
$ can be derived\footnote{The derivation of (\ref{P4}) is omitted due to space limitation} in closed-form as \ref{P4} on the top of next page \cite{zwillinger2014table},
\begin{figure*}[!t]
\vspace*{4pt}
\normalsize
\setcounter{MYtempeqncnt}{\value{equation}}
\begin{equation}\label{P4}
\Pr\left\{\left(\gamma_{SD}+\gamma_{RD}<\gamma_2\right)\cap\left(\gamma_{SD}<\gamma_1\right)\right\}
=F_{H_{SD}}\left(\frac{\gamma_1 N_0}{P_S}\right)-\sum^{N-1}_{p=0}\frac{\exp\left(-\frac{\gamma_2}{\bar{\gamma}_{RD}}\right)}{ \bar{\gamma}_{SD} \bar{\gamma}_{RD}^p p!}\sum_{k=0}^p \frac{\left( {\begin{array}{*{20}{c}}
p\\
k
\end{array}} \right)\gamma_2^{p-k}\left(-1\right)^k \Upsilon\left(k+1,\frac{\bar{\gamma}_{RD}-\bar{\gamma}_{SD}}{\bar{\gamma}_{SD}\bar{\gamma}_{RD}}\gamma_1\right)}{\left(\frac{\bar{\gamma}_{RD}-\bar{\gamma}_{SD}}{\bar{\gamma}_{SD}\bar{\gamma}_{RD}}\right)^{k+1}}
\end{equation}
\hrulefill
\vspace*{4pt}
\end{figure*}
where $\bar{\gamma}_{SD} = {P_S \Omega_{SD}}/{N_0}$ and $\bar{\gamma}_{RD} = {2 E_T \Omega_{RD}}/{N_0}$, and $\Upsilon\left(\alpha,x\right) = \int^x_0 \exp^{-t} t^{\alpha - 1} dt$ is the incomplete Gamma function. By substituting the terms derived above into (\ref{P_out}), we now have obtained a closed-form expression for the system outage probability of the proposed IATF protocol.

\section{Numerical Results}
\begin{figure}[!t]
  \centering\scalebox{0.5}{\includegraphics{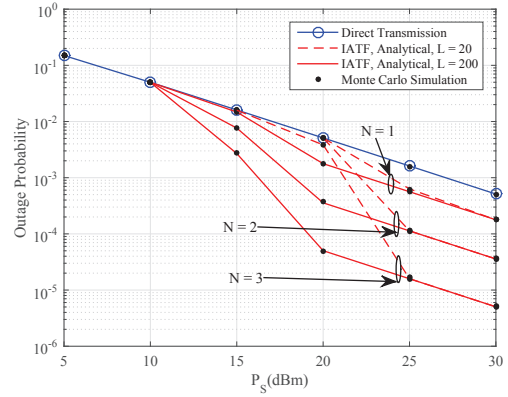}}
  \caption{ Outage probability of the proposed IATF scheme versus the source transmit power for different antenna numbers and battery levels, where $\mathbb{R}=1$, $C=5\times 10^{-3}$, $E_T=1\times 10^{-3}$, $N=[1,2,3]$.}\label{1}
\end{figure}
In this section, we provide some simulation results to verify the above theoretical analysis and illustrate the impacts of several parameters on system performance. We adopt the channel model $\Omega_{ij}=\left(1+d_{ij}^{\alpha}\right)^{-1}$ to capture the path-loss effect, where $d_{ij}$ denotes the distance between nodes $i$ and $j$, $\alpha\in\left[2,5\right]$ is the path-loss exponent. In the following simulations, we set $d_{SD}=80$m, $d_{SR}=10$m, $d_{RD}=70$m, the path-loss factor $\alpha=3$, the Rician-factor $K=10$, the noise power $N_0=-60$dBm, the energy conversion efficiency $\eta=0.5$, and the transmission rate of the system~$\mathbb{R}=1$.

\begin{figure}[!t]
  \centering\scalebox{0.5}{\includegraphics{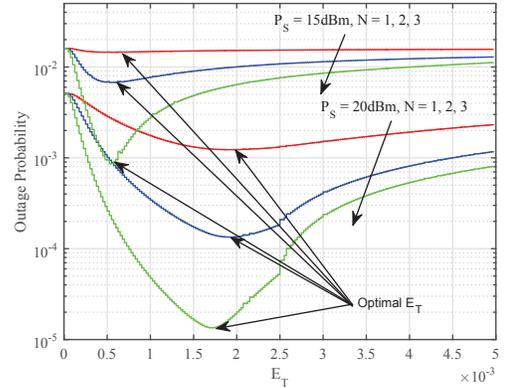}}
  \caption{ Outage probability of the proposed IATF scheme versus the consumed energy for information forwarding at the relay with different transmit power at the source, where $\mathbb{R}=1$, $C=5\times 10^{-3}$, $P_S=[15, 20]$dBm, $N = [1,2,3]$.}\label{2}
\end{figure}

The system outage probability of the proposed IATF scheme versus the source transmit power for different antenna numbers and battery levels is shown in Fig. \ref{1}. As we can see from this figure, Monte Carlo simulations agree very well with the corresponding analytical results plotted via (\ref{P_out}), which validates our theoretical analysis in Sec. III.
We can see from Fig. \ref{1} that, when $P_S$ is low, the outage probability of the proposed IATF scheme is similar to that of the direct transmission without cooperation. This is because the accumulated energy at the relay's battery cannot exceed the threshold. As $P_S$ increases, the gain of the proposed scheme over direct transmission becomes significant. We can also observe from this figure that when the number of battery levels increases from $20$ to $200$, the performance of the system gets better. This is because the more the battery being discretized (i.e., $L$ increases), the less energy is wasted during the discretization of the harvested energy. Furthermore, with more antennas equipped at the relay, the performance of the proposed IATF protocol improves.

Fig. \ref{2} plots the outage probability of the proposed IATF scheme versus energy threshold at the relay with different antenna numbers and source's transmit powers. This figure is shown as a stair-stepping plot due to the adopted discrete battery model. An optimal value of $E_T$ that minimizes the outage probability of the proposed scheme can be observed from the figure. We can also see that the values of optimal $E_T$ shift to the right as $P_S$ increases from $15$dBm to $20$dBm. This is because when $P_S$ increases, the direct transmission suffers from less outages and thus needs less help from $R$. In this case, $R$ can accumulate more energy in its battery and should consume more energy to compensate the received SNR at $D$. In addition, for a fixed $P_S$, the relay with more antennas requires a relatively smaller $E_T$.

\begin{figure}[!t]
  \centering\scalebox{0.5}{\includegraphics{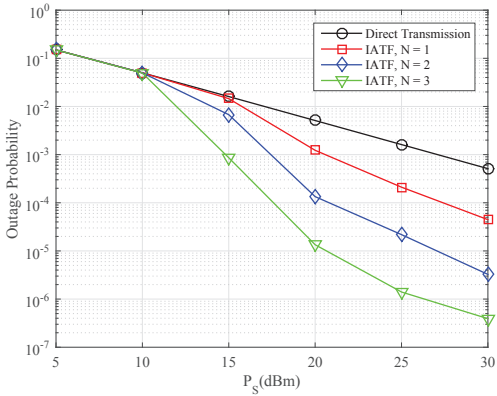}}
  \caption{ Outage probability of the proposed IATF scheme with optimal energy threshold at the relay, where $\mathbb{R}=1$, $C=5\times 10^{-3}$, $N=[1, 2, 3]$.}\label{3}
\end{figure}
For a given network setup, we can readily find the optimal value of $E_T$ via an exhaustive search of all discrete battery levels. The outage probability of the proposed IATF scheme with optimal $E_T$ is depicted in Fig. \ref{3}. We can observe that the proposed IATF protocol with optimal $E_T$ can significantly improve the outage performance compared to the direct transmission scheme, even when the relay is equipped with single antenna. Moreover, the performance gain increases as the number of antennas at relay increases. This is because the more antennas equipped at relay, the more energy it can harvest from the first hop. Moreover, with MRT, the received SNR at the destination can be larger.
\section{Conclusion}
In this paper, we proposed an incremental accumulate-then-forward (IATF) relaying protocol for cooperative communications via an EH relay. We modeled the energy accumulation process of the discrete-level relay battery as a finite-state Markov chain and derived a closed-form expression for the exact outage probability of the considered network over mixed Rician-Rayleigh fading channels. Monte Carlo simulation validated our theoretical analysis. Numerical results showed that the system outage probability decreases as either the source transmit power or number of antennas at the relay increases. Furthermore, the proposed IATF scheme can significantly outperform the direct transmission scheme without cooperation and their performance gap is further enlarged when the relay energy threshold is optimized.

\bibliographystyle{IEEEtran}
\bibliography{energy_harvesting_endnote}
\end{document}